\documentclass[aps,prb,twocolumn,floatfix,superscriptaddress]{revtex4}
\usepackage{graphicx}
\begin{document}

\title{Theory of highly excited semiconductor nanostructures including Auger coupling: exciton-bi-exciton mixing in CdSe nanocrystals }

\author{Marek Korkusinski}
\affiliation{Quantum Theory Group, Institute for Microstructural
  Sciences, National Research Council, Ottawa, Canada, K1A0R6}

\author{Oleksandr Voznyy}
\affiliation{Quantum Theory Group, Institute for Microstructural
  Sciences, National Research Council, Ottawa, Canada, K1A0R6}

\author{Pawel Hawrylak}
\affiliation{Quantum Theory Group, Institute for Microstructural
  Sciences, National Research Council, Ottawa, Canada, K1A0R6}

\begin{abstract}
We present a theory of highly excited interacting carriers confined in
a semiconductor nanostructure,
incorporating Auger coupling between excited states with different
number of excitations.  
The Coulomb matrix elements connecting exciton, bi-exciton and
tri-exciton complexes are derived and an intuitive picture of breaking
neutral multi-exction complexes into positively and negatively charged
multi-exciton complexes is given. 
The general approach is illustrated by analyzing the coupling of
biexciton and exciton in CdSe spherical nanocrystals.
The electron and hole states are computed using atomistic $sp^3d^5s^*$
tight binding Hamiltonian including an effective crystal field
splitting and surface passivation. 
For each number of electron-hole pairs the many-body spectrum is
computed in the configuration-interaction approach.
The low-energy correlated biexciton levels are broken into charged 
complexes: a hole and a negatively charged trion and 
an electron and  a positively charged trion. 
Out of a highly excited exciton spectrum a subspace coupled to
bi-exciton levels via Auger processes is identified.
The interaction between correlated bi-exciton and exciton states is
treated using exact diagonalization techniques. 
This allows to extract the spectral function of the biexciton and
relate its characteristic width and amplitude to the characteristic
amplitude and timescale of the coherent time evolution of the coupled
system. 
It is shown that this process can be described by the Fermi's Golden
Rule only if a fast relaxation of the excitonic subsystem is accounted
for.
\end{abstract}

\maketitle

\section{Introduction}
There is currently renewed interest in the understanding of
multi-exciton complexes in highly excited semiconductor nanostructures
and their interaction with light.  
Two examples are bi-exciton-exciton cascade for the generation of
entangled photon pairs in self-assembled quantum
dots\cite{stevenson_young_nature2006,akopian_lindner_prl2006,greilich_schwab_prb2006,reimer_korkusinski_prb2008,korkusinski_reimer_prb2009} 
and enhancing the efficiency of photovoltaic cells by  generation
of multi-exciton complexes (MEG) from a single, high-energy photon
absorbed in semiconductor nanocrystals (NCs).\cite{nozik_physE_2002,rabani_baer_nanolett08,schaller_pietryga_nanolett07,franceschetti_an_nanolett06,nair_bawendi_prb07,ellingson_beard_nanolett05,scholes_rumbles_natmat06}
While much progress has been achieved in the understanding of
multi-exciton complexes in self-assembled quantum
dots\cite{qdot_book,bimberg_book,michler-book} and 
nanocrystals,\cite{ekimov_hache_josa93,efros_rosen_prb96,leung_pokrant_prb98,chen_whaley_prb04,sapra_sarma_prb04,delerue_allan_prb05,wang_zunger_jpchem98,franceschetti_fu_prb99,califano_franceschetti_prb07,osovsky_cheskis_prl09,htoon_crooker_prl2009,furis_htoon_prb2006,sewall_franceschetti_prb09,achermann_hollingsworth_prb03,caruge_chan_prb04,bonati_mohamed_prb05,klimov_ivanov_nature07}
the mixing and decay of complexes with different numbers of excitons
are much less understood. 


These processes are important for the understanding of decoherence in
entangled photon pairs and energy-to-charge conversion.    
Nozik\cite{nozik_physE_2002} has put forward a proposal of converting the
energy of the  high-energy exciton generated by a high energy photon
into  several low-energy electron-hole pairs rather than allowing this
energy to be dissipated.
The theoretical threshold photon energy, at which the excited exciton
is expected to convert into a biexciton, depends on the nanocrystal
size, but is typically about twice the semiconductor bandgap, $2E_g$.
Such carrier multiplication process has been demonstrated in
PbSe, PbS, PbTe, CdSe, InAs, and Si 
NCs,\cite{schaller_klimov_prl2004,nozik_cpl2008} 
with efficiency reaching 700\% (seven electron-hole pairs out of one 
photon).\cite{schaller_sykora_nanolett2006}
However, a careful analysis of these experiments revised these
efficiencies to lower values.\cite{nair_bawendi_prb07,pandey_guyot_jcp2007,sewall_franceschetti_prb09}   

Coherent coupling of bi-exciton to highly excited exciton can lead to 
coherent conversion from biexciton to exciton and vice versa. 
However, phonon-assisted relaxation in the exciton subsystem was
suggested to destroy coherence and leads to finite bi-exciton
lifetime.\cite{nozik_arpc01} 
Exact theoretical estimation of this lifetime is challenging, as
even a realistic computation of single-particle states in NCs 
is difficult: for example, for a CdSe NC with diameter of $2.5$ nm
with $304$ atoms, in the energy range of  $3 E_g$ there are
approximately $1.58\cdot 10^5$ exciton and $6.2\cdot 10^9$ bi-exciton
states.  
As an approximation, one typically starts with the computation of NC
single-particle states using the $k\cdot p$~\cite{ekimov_hache_josa93,efros_rosen_prb96}, 
tight-binding\cite{leung_pokrant_prb98,chen_whaley_prb04,sapra_sarma_prb04,delerue_allan_prb05}
or empirical pseudopotential methods.\cite{wang_zunger_jpchem98,franceschetti_fu_prb99,califano_franceschetti_prb07} 
Next, one builds the electron-hole pair configurations to be coupled,
e.g., the exciton (X) and bi-exciton (XX) of similar energy, and
neglects all Coulomb coupling (correlations) within each subsystem.
Finally, one computes the Coulomb coupling matrix element between the
chosen X and XX configurations and computes the lifetime of the XX
state using the Fermi's Golden Rule.\cite{shabaev_efros_nanolett06,schaller_agranovich_natphys2005,allan_delerue_prb2006,allan_delerue_prb2008,rabani_baer_nanolett08,franceschetti_an_nanolett06}
As typically there are many X states close in energy to the XX state,
one utilizes a quasi-resonant approach, scaling the XX-X transition
rate by the density of X states and/or using a 
``resonance window''.\cite{allan_delerue_prb2006,allan_delerue_prb2008,rabani_baer_nanolett08,franceschetti_an_nanolett06,califano_acsnano2009}

A more advanced approach was presented recently by Witzel {\em et al.}\cite{witzel_shabaev_prl2010}
In this work, a time-dependent evolution of a single-photon excitation
coupled coherently with multi-exciton states, considered in the
$k\cdot p$ approach, is simulated.
Decay of each multiexciton state is accounted for in the
relaxation-time approximation.
It is shown that the relaxation times of different multiexciton
complexes, and in particular that of X, play a crucial role in the
efficiency of MEG. 

Engineering of materials and nanostructures directed at optimization
of the MEG gain requires therefore a comprehensive, microscopic
theory of (i) coupling of multiexciton states with different number of
excitations, and (ii) relaxation processes of these multiexciton
states.
In this work we focus on the former, as the latter requires
an additional realistic simulation of the phonon modes in the NC and a
treatment of the carrier-phonon coupling.\cite{kilina_kilin_acsnano2009,prezhdo_acr2009} 
We provide a detailed derivation of the Coulomb matrix element
coupling the multi-electron-hole configurations differing by one
electron-hole pair.\cite{hawrylak_wojs_prb96}
Further, we express the states of the system in the basis of
configurations with different number of excitations.
The form of this expansion is obtained by exact diagonalization of the
Hamiltonian accounting for {\em all} Coulomb interactions of quasiparticles.
From such an eigenstate we extract the spectral function of the state
with the larger number of pair excitations, assuming that it is
``immersed'' in a dense spectrum of the states with fewer pairs.
To make contact with the language of the state lifetime, used in
experiments, we relate the amplitude and the characteristic width of
this spectral function to the amplitude and time constants of the
coherent time evolution of the system.

We illustrate this framework on the mixing of X and XX in a spherical
CdSe nanocrystal.
The current work is built upon our recent calculation of the
electronic and optical properties of low energy X and XX states in the
CdSe NC.\cite{korkusinski_voznyy_prb2010}
We utilize the QNANO computational
platform\cite{korkusinski_zielinski_jap09} to carry out atomistic
tight-binding computation of the single-particle states in the NC
with diameter of $3.8$ nm. 
These states are used to construct the {\em correlated} ground state of the
XX as well as the excited states $X^*$ with energies close to that of
the XX.
From the exact diagonalization of the  XX interacting with $X^*$ in
this energy range we extract the spectral function of the XX ground
state and discuss its properties in connection with the coherent time
evolution of the coupled system without relaxation.
For the NC studied we find that the XX-$X^*$ coupling is  weak,
resulting in quantum beats between the XX and $X^*$ states.
We demonstrate that contact with the description of the XX population
in terms of lifetime can be established only if very fast decay
of the $X^*$ states due to phonons is assumed.

The paper is organized as follows. 
In Section II we present a framework theoretical approach to the
system in which the multiexciton states differing in the number of
excitations are coupled.
We establish the general form of the Hamiltonian for the system,
and, on an example of the $X-XX-XXX$ system,
demonstrate the derivation of the coupling Coulomb matrix elements.
We write the eigenstates of the coupled system, define the spectral
function of the system of $n$ electron-hole pairs immersed in the
spectrum of $n-1$ and $n+1$-pair excitations, and relate this function
to the time evolution of the system.
We illustrate these concepts in detail in Section III on the example
of the coupled $X-XX$ system confined in a single CdSe nanocrystal.
In Section IV we present conclusions and outlook.

\section{Model}

In this Section we present a general derivation of the Coulomb matrix
elements which couple the multiexciton configurations differing in the
number of electron-hole pairs.
We demonstrate how to include these elements in the exact
diagonalization study of the mixed system and how to extract
physically relevant quantities from that calculation.

\subsection{Derivation of the coupling Coulomb elements \label{model-coupling_elements}}

We start the derivation by writing the all-electron Hamiltonian for
the semiconductor nanostructure.
If by $c_i^+$ ($c_i$) we denote the creation (annihilation) operator
of an electron on state $|i\rangle$, we have:
\begin{equation}
  \hat{H} = \sum_i{\tilde{E}_i c^+_i c_i }
  + {1\over2} \sum_{ijkl} \langle i j | V_{ee} | kl \rangle
  c^+_i c^+_j c_k c_l,
  \label{hamiltonian_complete}
\end{equation}
where $\tilde{E}_i$ are the single-particle energies of
the nanostructure,
while $\langle i j | V_{ee} | kl \rangle$ are the Coulomb scattering
matrix elements computed for the single-particle states.
Since in the typical NCs we deal with $\sim 10^3$ atoms, or 
$\sim 10^4$ electrons, we cannot treat the above Hamiltonian directly
and introduce the language of quasiparticles.
To this end, we divide the basis of single-particle states into two
sets: the valence states, henceforth enumerated by Greek indices,
and the conduction states, enumerated with Latin indices.
As a result, the first term of the above Hamiltonian will split into
two, while the Coulomb operator will result in the appearance of
$2^4=16$ terms.
Among these terms there are those which describe the interaction of
carriers on conduction states only, as well as that of carriers
on valence states only.
Further, we find terms describing the interaction between conduction
and valence carriers, consisting of the direct and exchange component.
Next, we have terms which describe all possibilities of the Coulomb
scattering with transfer of one carrier from the valence to conduction
band or the other way around.
All of them consist of the ``direct''-like and ``exchange''-like
component. 
Finally, there are terms describing the Coulomb scattering with transfer
of {\em two} carriers, from the valence to conduction band 
and the other way around.
Due to the two-body character of Coulomb interactions, the above
Hamiltonian exhausts all possibilities of Coulomb scattering in the
system. 
We shall write all these terms explicitely below.

In order to complete the transition into the language of
quasiparticles, we define the ``vacuum'' reference state of our
system, in which all valence orbitals are occupied, and all conduction
orbitals are empty:
\begin{equation}
|0\rangle = \prod_\alpha{ c_{\alpha}^+ |vac\rangle},
\end{equation}
where $|vac\rangle$ denotes the true zero-electron state.
The energy of the state $|0\rangle$, $E_0$, is treated as the
reference level. 
In what follows, we will consider the charge-neutral electron-hole
excitations from that state, which can be written as
\begin{equation}
|i,j,k,\ldots,\alpha,\beta,\gamma,\ldots\rangle =
c_i^+ c_j^+ c_k^+\ldots h_{\alpha}^+ h_{\beta}^+ h_{\gamma}^+ \ldots
|0\rangle. 
\label{ehconfig}
\end{equation}
Here the hole creation (annihilation) operators are defined as
$h_{\alpha}^+ = c_{\alpha}$ ($h_{\alpha} = c_{\alpha}^+$),
respectively.
As is evident from the discussion opening this Section, 
the Hamiltonian (\ref{hamiltonian_complete}) describes direct coupling
of a configuration with $n$ electron-hole pairs and configurations
with $n-2$, $n-1$, $n$, $n+1$, and $n+2$ pairs.
However, we need to translate it into the language of quasiparticle
operators.
By replacing the valence operators with the hole operators and
rearranging terms, we obtain:
\begin{equation}
\hat{H}_{QP} = \hat{H}_{CONS} + \hat{H}_{NC},
\label{hamiltonian_quasiparticle}
\end{equation}
where the part conserving the number of excitations is
\begin{eqnarray}
  \hat{H}_{CONS} &=& \sum_i{E_i c^+_i c_i } 
  + {1\over2} \sum_{ijkl} \langle i j | V_{ee} | kl \rangle
  c^+_i c^+_j c_k c_l \nonumber\\
&-&\sum_{\alpha}{E_{\alpha} h^+_{\alpha} h_{\alpha} } 
  + {1\over2} \sum_{\alpha\beta\gamma\delta} 
  \langle \delta \gamma | V_{ee} | \beta \alpha \rangle
  h^+_{\alpha} h^+_{\beta} h_{\gamma} h_{\delta} \nonumber\\
&-& \sum_{i\beta\gamma l} 
\left(\langle i \gamma | V_{ee} | \beta l \rangle
- \langle i \gamma | V_{ee} | l \beta  \rangle \right)
  c^+_{i} h^+_{\beta} h_{\gamma} c_{l} ,
\end{eqnarray}
and the part changing the number of excitations is
\begin{eqnarray}
  \hat{H}_{NC} &=& {1\over2} \sum_{i j k \delta} 
  \left(\langle i j| V_{ee} | k \delta \rangle
    -  \langle i j| V_{ee} | \delta k \rangle\right)
  c^+_{i} c^+_{j} c_{k} h_{\delta}^+ \nonumber \\ 
  &+& {1\over2} \sum_{i \beta k l} 
  \left(\langle i \beta | V_{ee} | k l \rangle
    -  \langle \beta j| V_{ee} | k l \rangle\right)
  c^+_{i} h_{\beta} c_{k} c_{l} \nonumber \\ 
  &+& {1\over2} \sum_{\alpha\beta\gamma l} 
  \left(\langle \alpha\beta| V_{ee} | \gamma l \rangle
    -  \langle \alpha\beta | V_{ee} | l \gamma \rangle\right)
  h_{\alpha} h_{\beta} h_{\gamma}^+ c_{l} \nonumber \\ 
  &+& {1\over2} \sum_{\alpha j \gamma \delta} 
  \left(\langle \alpha j | V_{ee} | \gamma \delta \rangle
    -  \langle j \alpha | V_{ee} | \gamma\delta \rangle\right)
  h_{\alpha} c^+_{j} h^+_{\gamma} h^+_{\delta} \nonumber \\ 
  &+& {1\over2} \sum_{i j\gamma\delta} 
  \langle i j | V_{ee} | \gamma \delta \rangle
  c^+_{i} c^+_{j} h^+_{\gamma} h^+_{\delta} \nonumber\\
  &+& {1\over2}\sum_{\alpha\beta k l} 
  \langle \alpha \beta | V_{ee} | k l \rangle
  h_{\alpha} h_{\beta} c_{k} c_{l}.
\end{eqnarray}
In the above Hamiltonian, the single-particle energies are those of
quasiparticles, and therefore have to be properly dressed in
selfenergy and vertex correction terms.
We shall not analyze them in greater detail, as all methods of
calculation of the single-particle structure, i.e., $k\cdot p$,
tight-binding, or pseudopotential approaches, at some point are fitted
to the experimental bandgaps, and therefore are parametrised with
already dressed single-particle energies.
Note that all Coulomb elements are computed with previous, electron
rather than the quasiparticle orbitals.
These orbitals are computed directly in the single-particle methods
and care must be taken to translate the Coulomb elements into the
quasiparticle language.
For example, the hole Coulomb matrix element
$\langle \alpha\beta|V_{hh}|\gamma\delta\rangle
=\langle \delta\gamma|V_{ee}|\beta\alpha\rangle$, that is, it is a
complex conjugate of the electron element, while the relation for
electron-hole interactions is even more complicated.

The form of the Hamiltonian (\ref{hamiltonian_quasiparticle}) allows
to appreciate the terms changing the number of excitations in a
clearer fashion: the terms changing the number of excitations by one
have three quasielectron and one hole operators (or conversely), while
the terms changing the number of excitations by two are built of four
creation or four annihilation operators.

\begin{figure}
\includegraphics[width=0.4\textwidth]{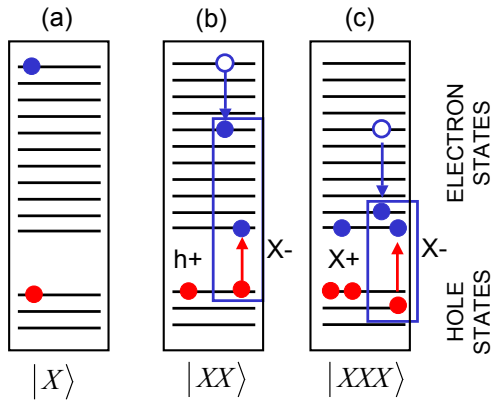}
\caption{ (Color online)
  Schematic diagram of the one-pair excitation (a),
  the two-pair excitation (b), and the three-pair excitation (c).
  Empty circles denote removed quasiparticles.
  The negatively charged trion $X-$ appearing as a result of this
  scattering event is marked with the rectangles.}
\label{fig0}
\end{figure}

In semiconductor nanostructures the scattering with transfer of one
carrier must necessarily modify the energy of a configuration by at
least one gap energy $E_g$, while the transfer of two carriers
introduces a modification of at least $2E_g$.
The largest mixing of configurations with different numbers of
excitations will be seen for configurations close in energy.
Therefore, in order to obtain coupling between a low-lying two-pair
excitation, or bi-exciton (XX), and an exciton (X), one has to
consider highly excited exciton states (excited by at least
$E_g$).\cite{nozik_physE_2002} 
The X-XX-tri-exciton mixing will start at even higher energies (of at
least $3E_g$), i.e., both X and XX must be highly excited.
We shall describe this case here in a greater detail.
Let us first write explicitely the Coulomb matrix elements
coupling the X and XX.
We couple a one-electron-hole-pair configuration
$|X,i\alpha\rangle=c_i^+h_{\alpha}^+|0\rangle$ 
with a two-pair configuration 
$|XX,jk\beta\gamma\rangle=c_j^+c_k^+h_{\beta}^+h_{\gamma}^+|0\rangle$ 
and obtain:\cite{hawrylak_wojs_prb96}
\begin{eqnarray}
\langle XX,jk\beta\gamma| &\hat{H}_{QP}& | X,i\alpha\rangle \nonumber\\
&=& \left[ 
\left( \langle jk|V_{ee}|\beta i \rangle - \langle jk|V_{ee}|i
  \beta\rangle \right) \delta_{\alpha\gamma} \right.\nonumber\\
&+&\left.
\left( \langle jk|V_{ee}|i \gamma  \rangle - \langle jk|V_{ee}|\gamma i
\rangle \right) \delta_{\alpha\beta} \right]\nonumber\\
&+& \left[ 
\left( \langle \alpha j|V_{ee}|\gamma\beta \rangle - \langle \alpha j|V_{ee}|
  \beta\gamma\rangle \right) \delta_{ik} \right.\nonumber\\
&+&\left.
\left( \langle \alpha k|V_{ee}|\beta \gamma  \rangle - \langle \alpha k|V_{ee}|\gamma \beta
\rangle \right) \delta_{ij} \right].
\label{mixing_element}
\end{eqnarray}
In the above formula, the terms in the first square bracket apply to
the scattering event whereby the creation of an electron-hole pair is
accompanied by a scattering of the electron constituting the exciton,
while the exciton's hole {\em must not} change its orbital.
The second square bracket contains analogous terms for the case when
the exciton's hole is scattered, but its electron must stay on the
same orbital.
This gives a selection rule for the pair creation process: the exciton
and biexciton configurations must share at least one carrier.
The process in which the hole is shared is illustrated in the transition
Fig.~\ref{fig0}(a)-(b).
In Fig.~\ref{fig0}(a) we show schematically a one-pair configuration,
in which the hole occupies the single-particle ground state, while the
electron is highly excited.
In the scattering process, the electron transfers to a lower
single-particle level, while an additional electron-hole pair is
created.
In Fig.~\ref{fig0}(b) this negatively charged trion complex $X-$ is
marked with the rectangle, with the empty circle 
denoting the initial level of the excited electron.
Note that the shared hole is a spectator quasiparticle and does not
take part in this process.
The scattering event with shared electron would conform to similar
rules, only in this case we would deal with a highly excited hole
converting to a positively charged trion, while the shared electron
would be inert.
In the case of the X-XX coupling, the selection rule severely limits
the number of excited X configurations to which any XX configuration
can couple directly. 
Henceforth we shall refer to these exciton configurations as $X^*_1$.
However, there are also excited X configurations, which are not
directly coupled to the XX configuration, but do couple to $X^*_1$.
These configurations shall be referred to as $X^*_2$.
We shall discuss the importance of $X^*_2$ later on.

Let us now move on to the coupling between the two- and three-pair
excitations, whose one variant is visualized schematically in
the transition Fig.~\ref{fig0}(b)-(c). 
The matrix element between the two-pair configuration
$|XX,jk\beta\gamma\rangle=c_j^+c_k^+h_{\beta}^+h_{\gamma}^+|0\rangle$
and the three-pair configuration
$|XXX,lmn\delta\sigma\pi\rangle=c_l^+c_m^+c^+_nh_{\delta}^+h_{\sigma}^+h_{\pi}^+|0\rangle$
is computed similarly to the one connecting the one- and two-pair
configurations.
We shall not write it here as it is composed of a large number of
terms describing various permutations of particles.
Let us only emphasize that the scattering process involves a creation
of a negatively charged trion $X-$ out of one excited electron, or a
positively charged trion out of one excited hole, just as in the case
of the X-XX coupling.
Figure~\ref{fig0}(c) depicts the former, as the excited electron is
scattered down the ladder of single-particle states (blue arrow) with
a simultaneous creation of the third electron-hole pair (red arrow).
The resulting negatively charged trion is denoted in the right-hand
panel of Fig.~\ref{fig0}(b)  by the blue rectangle.
The remaining three ``spectator'' particles are inert
and form a positively charged trion $X+$.

Finally, there is also a possibility of coupling the one- and three-pair
exictations directly via the two last terms of the Hamiltonian
(\ref{hamiltonian_quasiparticle}).
However, in this case the energies of the configurations must
necessarily differ by at least $2E_g$.
This is why in the following we shall neglect this term.

\subsection{Eigenstates of the mixed system\label{eigenstates_mixed}}

Presently we solve for the eigenvalues and eigenstates of the system
with a non-constant number of excitations.
To this end, we use a procedure consisting of two steps.
First, we solve for eigenstates and eigenenergies of the Hamiltonian
(\ref{hamiltonian_quasiparticle}) in the subspaces spanned by the
configurations with a {\em conserved} number of excitations.
In this case, only the first five terms of that Hamiltonian are
considered.
We solve this problem in the exact diagonalization approach, in which
the Hamiltonian matrix is written in the basis of configurations of
the type given in Eq.~(\ref{ehconfig}) and diagonalized numerically.
As a result of this procedure, for a system with $n$ electron-hole
pairs we obtain the eigenstates of the form 
\begin{equation}
|nX\rangle_p = \sum_{i,j,\ldots,\alpha\beta\ldots}
A^{p,n}_{i,j,\ldots,\alpha\beta\ldots} |i,j,\ldots,\alpha,\beta\ldots\rangle,
\label{many_exciton_state}
\end{equation}
where the coefficients $A^{p,n}_{i,j,\ldots,\alpha,\beta\ldots}$ compose
the $p$-th eigenvector of the Hamiltonian matrix, and the energy of
this state is $E_{p,n}$.
Since the number of possible $n$-pair configurations can be very
large, we restrict the basis to the region of energies of interest and
control the convergence of the resulting energy levels with the width
of that region.

In the second step, utilizing the energies and eigenstates of systems
with conserved excitation numbers, we set up the full matrix of the
Hamiltonian (\ref{hamiltonian_quasiparticle}).
The only nondiagonal elements in this matrix will result from the
Hamiltonian terms changing the number of excitations.
They are computed as linear combinations, which, for example, for the
coupled $X^*_1$-XX system take the form:
\begin{eqnarray}
&\left._p\langle\right.& XX| \hat{H}_{QP} |X^*_1\rangle_q \nonumber\\
&=& \sum_{j,k,\beta,\gamma} \sum_{i,\alpha}
\left( A^{p,XX}_{k,l,\beta,\gamma} \right)^*
A^{q,X}_{i,\alpha}    
\langle k l \beta \gamma | \hat{H}_{QP} | i, \alpha \rangle.
\label{coupling_correlated}
\end{eqnarray}
Note that although the individual coupling elements under the sum may
vanish due to the selection rule described above, the elements
connecting the correlated states may still be finite.
We diagonalize the Hamiltonian matrix to obtain the eigenstates with
mixed number of excitations.
The $K$-th state can be written as the linear combination:
\begin{equation}
|K\rangle = B^K |0\rangle
+ \sum_{i} \sum_n C^K_{i,n} |nX\rangle_i.
\label{mixed_state}
\end{equation}
The energy of this state is referred to as $\varepsilon_K$.

\subsection{Spectral function and its relation to time evolution\label{spectral_theory}}

The degree of mixing between states with different number of
excitations can be extracted from the eigenstates $|K\rangle$ by
calculating the spectral function $A_{p,n}(K)$ of the
$p$-th state with $n$ electron-hole pairs $|nX\rangle_p$.
This function is computed as
\begin{equation}
A_{p,n}(K) = |\left._p\langle\right. nX|K\rangle|^2 
\end{equation}
and, for example, for the $p$-th biexciton state will take the form
$ A_{p,2}(K) = | C^K_{p,2} |^2 $, i.e., it is readily obtained from the
eigenvectors of the system.
For weak coupling we expect that the spectral function will have the
value close to $1$ for $K$ such that $\varepsilon_K\approx E_{p,n}$,
and decay as we move to the other eigenstates $|K\rangle$.

In practice, one is typically interested in the lifetime of the
multiexciton state and attempts to engineer the system so as to
achieve long lifetimes of states with many excitons.
If the dynamics of the system is governed by the Hamiltonian
(\ref{hamiltonian_quasiparticle}) only and is not changed by any
incoherent relaxation processes, we can trace the time evolution of
the state $|\Psi\rangle$ of the coupled system simply by
\begin{equation}
|\Psi(t)\rangle = \exp\left( -{i\over\hbar} \hat{H}_{QP} t\right) |nX\rangle_p,
\end{equation}
assuming that the system is prepared in the state $|nX\rangle_p$.
Since $|nX\rangle_p = \sum_K \left( C^K_{p,n} \right)^* |K\rangle$,
we have 
\begin{equation}
|\Psi(t)\rangle = 
\sum_K \exp\left( -{i\over\hbar} \varepsilon_K t\right) 
\left( C^K_{p,n} \right)^* |K\rangle,
\end{equation}
and we can observe the time evolution of our state by computing the
projection
$|\left._p\langle\right. nX|\Psi(t)\rangle|^2
= \left| \sum_K \exp\left( -{i\over\hbar} \varepsilon_K t\right) 
\left| C^K_{p,n} \right|^2  \right|^2
= \left| \sum_K \exp\left( -{i\over\hbar} \varepsilon_K t\right) 
A_{p,n}(K)  \right|^2$.
The time evolution of the system, and in particular the change of the
number of excitations from the one prepared in the system, is related
to the Fourier transform of the spectral function in the time
domain.
At this point it is useful to consider two limiting examples of the
spectral function.
First, if $A_{p,n}(K)$ is finite only for several states $K$,
we may expect a complex time evolution, with oscillating contributions
from all these states.

Second, if our state $|nX\rangle_p$ is immersed in a quasi-continuous
spectrum of other states (possibly with different $n$) and if
its spectral function can be approximated by a Lorentzian,
$A_{p,n}(K)\approx (\gamma/2\pi) / [ ( \varepsilon_K - E_{p,n} )^2 +
(\gamma/2)^2 ]$ and $A_{p,n}(K)= A_0$ on resonance, then the time
evolution is described by an exponential decay,
$|\left._p\langle\right. nX|\Psi(t)\rangle|^2
= |A_0 + (1-A_0) \exp ( - \gamma t / 2)|^2$,
with $\gamma$ being the characteristic width of the spectral function.
At long times and with strong mixing (the value of $A_0 \ll 1$), the
state of the system can no longer be identified with the state
$|nX\rangle_p$, as the probability density is 
distributed {\em coherently} in the multitude of states of the system.
The above analysis shows that the characteristic decay time constant
$1/\gamma$ is {\em not} related simply to the bare coupling between
the state $|nX\rangle_p$ and other states {\em at the same energy}, as
this will predominantly affect the spectral function maximum value
$A_0$. 
The dynamics of this ``dissolution'' of our state $|nX\rangle_p$
is decided by its coupling to states off-resonance: the stronger that
coupling, the broader the spectral function and the faster the decay.
Note also that the probability of finding the system in the state 
$|nX\rangle_p$ does not decay to zero, but rather the amplitude
$A_0^2$. 
We shall supplement this intuitive picture with a detailed analysis of
the dynamics in the next Section.

\section{Exciton-biexciton coupling in CdSe nanocrystals}

In this Section we apply the general approach to describing the
dynamics of the bi-exciton XX in a CdSe nanocrystal with diameter of
$3.8$ nm, as studied in Ref.~\onlinecite{korkusinski_voznyy_prb2010}.
We choose to focus on the low-energy XX states, which allows us to
consider their coupling only to excited exciton states of similar
energy.
Below we will discuss the entire procedure, starting from the
computation of single-particle states, then the correlated states of X
and XX, coupling matrix elements, treatment of the mixed system and
extraction and analysis of the spectral function.

\subsection{Single-particle states}

We start the parametrization of the Hamiltonian
(\ref{hamiltonian_quasiparticle}) with the computation of
single-particle states and their energies.
To this end we utilize the atomistic tight-binding $sp^3d^5s*$
approach.
We look for the single-particle wave function $|i\rangle$ in the form
of a linear combination
\begin{equation}
|i\rangle = \sum_{R,\alpha} F^{(i)}_{R,\alpha}|R,\alpha\rangle
\end{equation}
of atomic orbitals $|R,\alpha\rangle$.
The index $\alpha$ enumerates the types of orbitals ($s$, three $p$,
five $d$, and $s*$, of which all are degenerate spin-doublets), while
the index $R$ enumerates atoms.
The coefficients $F^{(i)}_{R,\alpha}$ and the energies $E_i$ of states
are computed by diagonalizing the tight-binding Hamiltonian
\begin{eqnarray}
\hat{H}_{TB} &=& 
\sum_{R,\alpha} \varepsilon_{R,\alpha} c^+_{R,\alpha} c_{R,\alpha}
+ \sum_{R,\alpha,\beta} \lambda^{SO}_{R,\alpha,\beta} c^+_{R,\alpha}
c_{R,\beta} \nonumber \\
&+& \sum_{R,\alpha} \sum_{R',\beta}
t^{R',\beta}_{R,\alpha} c^+_{R,\alpha} c_{R',\beta}
\label{hamiltonian_tb}
\end{eqnarray}
written in the basis of atomic orbitals.
This Hamiltonian is parametrized by the on-site energies
$\varepsilon_{R,\alpha}$, spin-orbit parameters
$\lambda^{SO}_{R,\alpha,\beta}$, and hopping elements
$t^{R',\beta}_{R,\alpha} c^+_{R,\alpha}$.
These parameters are established by fitting the bulk band structure
obtained with the above Hamiltonian to the structure obtained either 
with {\em ab initio} methods or
experimentally.\cite{korkusinski_voznyy_prb2010}
In Ref.~\onlinecite{korkusinski_voznyy_prb2010} we have presented an
extensive analysis of the electronic and optical properties of NCs
using the atomistic tight-binding method.
Here we will summarize it briefly.

We focus on the spherical CdSe nanocrystal with diameter of $3.8$ nm.
The underlying crystal lattice of the NC is of the wurtzite
modification.
The lattice symmetry induces the crystal field, whose effects are
apparent only at the third-nearest neighbor distances, and therefore
are not treated naturally with the nearest-neighbor Hamiltonian
(\ref{hamiltonian_tb}). 
We model the crystal field by introducing a splitting in energies of
the $p$ orbitals on each atom.
The NC surface is passivated in a model approach by applying a large
energy shift to any dangling bond.\cite{lee_oyafuso_prb03}
The sample is composed of $10^3$ atoms, which results in the
tight-binding Hamiltonian matrix of $2\cdot 10^4$.
Since in our further analysis we require single-particle states whose
energies fall into a large region (approximately from $-2E_g$ to
$2E_g$), we perform the diagonalization of our Hamiltonian using
full-matrix diagonalization tools.

\begin{figure}
\includegraphics[width=0.4\textwidth]{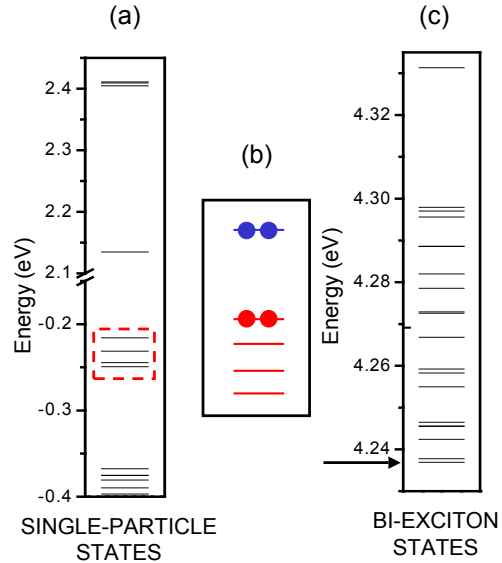}
\caption{(Color online)
  (a) Energies of several lowest-lying single-particle
  electron and hole states of a CdSe spherical
  nanocrystal with the diameter of $3.8$ nm.
  Bars denote Kramers doublets.
  (b) Schematic representation of the dominant two-pair configuration
  of the lowest bi-exciton state. Arrows distinguish the states within
  each Kramers doublet.
  (c) Lowest-energy segment of the bi-exciton spectrum.
  Arrow denotes the lowest XX state.
}
\label{fig1}
\end{figure}

The energies of several lowest-lying electron and hole states
are visualized in Fig.~\ref{fig1}(a).
Here, bars represent the energies of Kramers doublets, which are
degenerate due to the time-reversal symmetry.
The electron states correspond well with those of a spherical quantum
well, with a single $s$-like level and three quasi-degenerate $p$-like
levels about $270$ meV higher.
The hole states, on the other hand, form a low-energy,
quasi-degenerate shell composed of four Kramers doublets, separated
from the rest of the spectrum by a gap of about $120$ meV.
The complicated structure of the hole states is a consequence of the
spin-orbit interactions and the presence of the crystal field.

In order to complete the construction of the quasiparticle Hamiltonian
(\ref{hamiltonian_quasiparticle}), we need to compute all relevant
Coulomb matrix elements using the single-particle TB wave functions. 
We use the expansion, in which we separate the on-site terms arising
from the scattered particles residing on the same atom and
the long-distance terms describing scattering between more
remote atoms.
\begin{eqnarray}
{\left\langle ij \right|} V_{ee} {\left| kl \right\rangle}
&=& V_{ons} + V_{long}, 
\label{coulomb_elements}\\
V_{ons} &=& \sum_{R}{
\sum_{\alpha\beta\gamma\delta}{
F_{R,\alpha}^{(i)*}
F_{R,\beta}^{(j)*}F_{R,\gamma}^{(k)}
F_{R,\delta}^{(l)} }} \nonumber\\
&\times& 
\langle\left. R,\alpha,R,\beta \right|
{e^2\over\epsilon_{ons}|\vec{r}_1-\vec{r}_2|}
\left| R,\gamma, R,\delta\right.\rangle, \\
V_{long} &=& \sum_{R_i}{\sum_{R_j}^{remote}{
\sum_{\alpha\beta}{
 F_{R_i,\alpha}^{(i)*}
 F_{R_j,\beta}^{(j)*} F_{R_j,\beta}^{(k)}
F_{R_i,\alpha}^{(l)}}}} \nonumber\\
&\times& 
{e^2\over\epsilon_{long}|\vec{R}_i-\vec{R}_j|}. 
\label{coulomb_remote}
\end{eqnarray}
The integrals scaling the onsite terms are
computed by approximating the atomistic functions $|R,\alpha\rangle$
by Slater orbitals.\cite{slater-pr1930}
In an attempt to simulate the distance-dependent dielectric
function,\cite{franceschetti_fu_prb99,wang_califano_prl03,moreels_allan_prb10,ogut_burdick_prl03,delerue_lannoo_prb03}
each of these terms is scaled by a different dielectric constant
$\epsilon$.
Typically we take $\epsilon_{ons}=1$.
As for the long-distance term, if the two atoms in question are
nearest neighbors, we take $\epsilon_{long}=2.9$, while for more
remote pairs we take $\epsilon_{long}=5.8$, the
latter one being the bulk CdSe dielectric constant.

\subsection{Biexciton and excited exciton}

Next we populate the single-particle states with electron-hole pairs.
Since the energy range of interest for the exciton-bi-exciton coupling
is defined by the energies of XX states, we start with considering the
two-electron-hole-pair configurations
\begin{equation}
|i,j,\alpha,\beta\rangle = c^+_i c^+_j h^+_{\alpha} h^+_{\beta} |0\rangle.
\end{equation}
We stress that the quasihole single-particle wave functions are
obtained by applying a complex conjugate to the valence functions
obtained from diagonalization of the tight-binding Hamiltonian.
We build the matrix of the many-body Hamiltonian
(\ref{hamiltonian_quasiparticle}) in the basis of these configurations
and diagonalize it numerically.
We have presented an extensive study of the electronic and optical
properties of XX in Ref.~\onlinecite{korkusinski_voznyy_prb2010}.
We found that the XX eigenstates have a correlated character, which is
due to the gaps between the single-particle hole states being of the
same order as the Coulomb scattering matrix elements between the two-pair
configurations (tens of meV).
By constraining the two electrons to occupy the $s$-shell electron
orbital and distributing the holes among the 4
double-degenerate states of the hole shell [marked in
Fig.~\ref{fig1}(a) with a red rectangle] we obtain $28$ two-pair
configurations which mix with each other strongly.
The configuration of the four carriers with the lowest single-particle
energy is shown schematically in Fig.~\ref{fig1}(b).
However, the basis restricted to these configurations only is not
sufficient to obtain convergence of the XX energies.
A satisfactory spectrum of XX energies, shown in Fig.\ref{fig1}(c), is
obtained only when the electrons are allowed to populate the $s$- and
$p$-shell single-particle orbitals, while holes are allowed to spread
on $14$ lowest-energy Kramers doublets.
This results in $10976$ two-pair configurations.
Inclusion of the  mixing between the two-pair configurations decreases
the XX ground-state energy to the value of $E_{1,2}=4.237$ eV, that
is, by about $75$ meV with respect to the uncorrelated case.

In Fig.~\ref{fig1}(c) we show only the  XX correlated states
built predominantly from the original $28$ two-pair configurations.
The respective XX eigenstates are linear combinations of all
basis states, represented in general by Eq. (\ref{many_exciton_state})
and in this case taking the form
\begin{equation}
|XX\rangle_p = \sum_{i,j,\alpha,\beta}
A^{p,XX}_{i,j,\alpha,\beta} |i,j,\alpha,\beta\rangle.
\label{correlated_biexciton}
\end{equation}
However, these states are typically dominated by one configuration.
In Fig.~\ref{fig1}(b) we show the two-pair configuration dominant in
the lowest-energy XX state.
The segment of the spectrum built upon the lowest-shell configurations
is separated from the rest of XX states by a gap resulting from the
gap in the hole single-particle spectrum.
This defines a region of energies appropriate for the studies of X-XX
mixing. 
In what follows, we shall construct the excited exciton
configurations, whose energy falls within the window of $4.0$ to
$4.35$ eV, that is, that corresponding to the lowest XX band enlarged
from each side by about $0.25$ eV.

\begin{figure}
\includegraphics[width=0.4\textwidth]{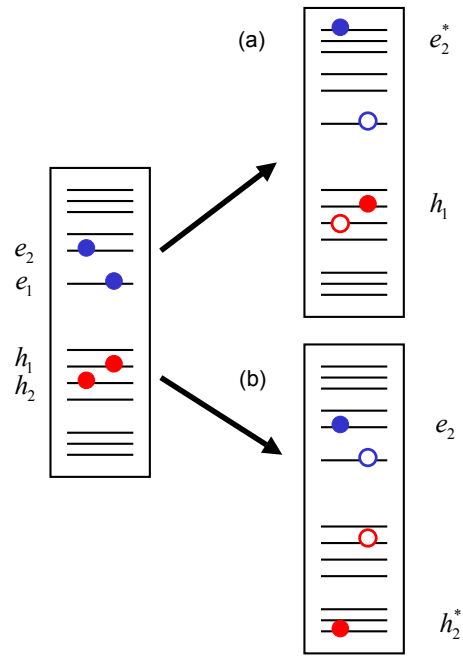}
\caption{(Color online)
  Schematic representation of the allowed Coulomb scattering
  mechanism connecting the two-pair and one-pair configurations.
  Empty circles denote removed quasiparticles.
  Panels (a) and (b) show respectively the event in which the
  electron and hole are scattered.}
\label{fig3}
\end{figure}

Let us now describe the procedure of generation of the excited X
states. 
As was explained in Section \ref{model-coupling_elements}, in order
to be able to couple, a two-pair configuration and an excited one-pair
configuration have to share at least one carrier.
Figure~\ref{fig3} shows how such $X_1^*$ configurations are created
assuming that they share a hole (a) or an electron (b) with a
two-pair configuration.
In this way, from each two-pair configuration one can generate
families of $X^*_1$ one-pair configurations and select only those,
whose energy falls in the chosen window, as described above.
However, generating all eligible $X^*_1$ from all $10976$ two-pair
configurations would result in a prohibitive size of the $X^*_1$
space.
Therefore, we only generate the $X^*_1$ families from the $28$
lowest-shell configurations.

\begin{figure}
\includegraphics[width=0.4\textwidth]{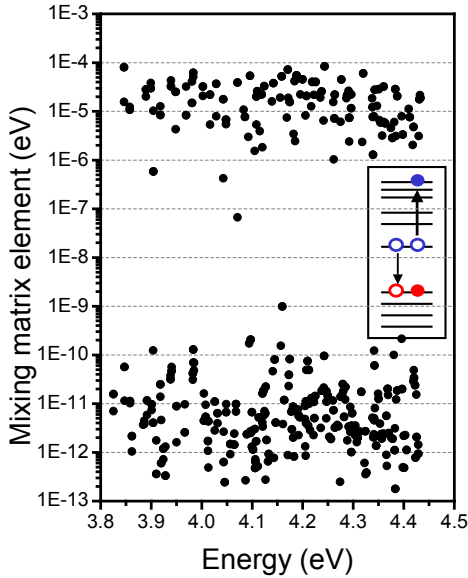}
\caption{(Color online)
  Magnitudes of the Hamiltonian matrix element coupling the
lowest-energy bi-exciton configuration [shown in Fig.~\ref{fig1}(b)]
with single-pair configurations with excited electron, as a function
of the energy of these configurations.
Inset shows a schematic illustration of the scattering process.}
\label{fig4}
\end{figure}

\begin{figure}
\includegraphics[width=0.4\textwidth]{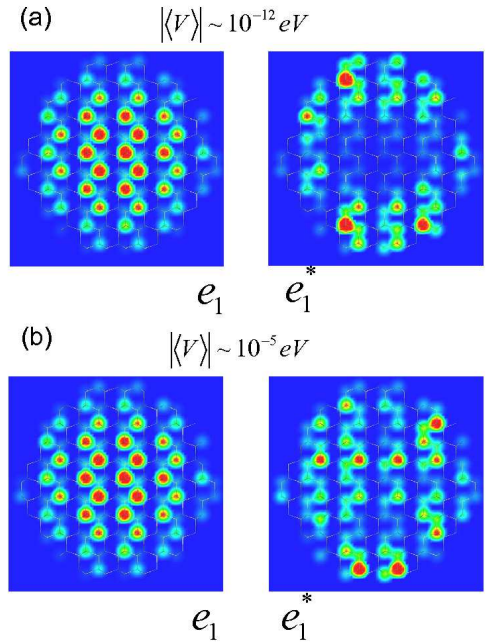}
\caption{(Color online)
  Cross-sectional view of the probability density of
  electronic states taking part in the scattering process depicted in
  Fig.~\ref{fig4}.
  The left-hand images correspond to the electron wave function on the
  $s$ orbital, while the right-hand images characterize the wave
  functions of the final state.
  Due to the distribution of the probability density, the element
  corresponding to the situation (a) is negligible, while that
  corresponding to the situation (b) is finite.}
\label{fig5}
\end{figure}

The next step is to compute all relevant Coulomb matrix elements
between configurations as defined by Eq.~(\ref{mixing_element}).
As an example, in Fig.~\ref{fig4} we show the absolute values of
elements connecting the lowest-energy two-pair configuration with
the $X^*_1$ configurations which share a common hole
(see a schematic diagram in the inset to this Figure).
We see that the elements fall into two categories.
Some elements have values ranging from $10^{-6}$ eV to $2\cdot
10^{-4}$ eV, while other elements are of $10^{-9}$ eV and less.
We find a similar distribution of the values of matrix elements for
$X_1^*$ configurations which share a common electron with the two-pair
configurations. 
The elements from the second category are within the numerical noise
of our computation and should be treated as zero, although they
quantify an allowed coupling process.
The source of such a large disparity of magnitudes of these elements
lies in the character of the wave function of the excited electron,
promoted to a high orbital due to the scattering.
In Fig.~\ref{fig5} we show the single-particle probability densities
of the orbitals which participate in the scattering.
The electron is promoted from the $s$-type orbital (left) onto one of
the highly excited orbitals (right).
In the case (a) we find that the final orbital is distributed in an
outermost layer of fully-coordinated Cd atoms of
the nanocrystal such that the overlap with the tightly confined $s$
orbital is small. 
This results in a negligible coupling.
On the other hand, in the case (b) the overlap between the relevant
orbitals appears to be more substantial, resulting in a much larger 
coupling of $~10^{-5}$ eV.
Note that all these matrix elements are at least two orders of
magnitude smaller than those connecting different two-pair
configurations to each other.\cite{korkusinski_voznyy_prb2010}

\subsection{Spectral function and dynamics of XX}

We now proceed to computing the electronic properties and dynamics of
the coupled XX-X system.
According to the procedure described in
Section~\ref{eigenstates_mixed}, having diagonalized the XX subsystem
separately, we should now perform a separate diagonalization of the X
subsystem and then connect the two.
We found, however, that a more computationally efficient procedure
involves writing the entire Hamiltonian
(\ref{hamiltonian_quasiparticle}) in the basis of the single-pair and
two-pair configurations.
This is due to the fact that in such large bases, the calculation of
elements (\ref{coupling_correlated}) coupling correlated $X^*_1$ and
XX states takes a very long time.

\begin{figure}
\includegraphics[width=0.4\textwidth]{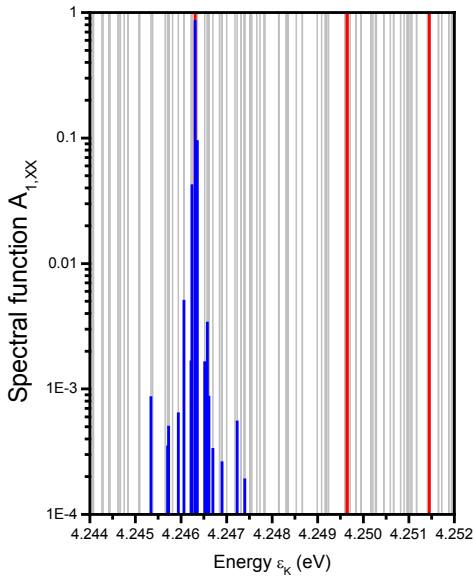}
\caption{(Color online)
 Spectral function of the bi-exciton immersed in the spectrum
  of exciton states. Red lines indicate the energies of the bi-exciton
  eigenstates, while gray lines correspond to the exciton eigenstates.
  Blue bars represent the values of the spectral function of the
  lowest-energy bi-exciton for each eigenstate of the coupled system.} 
\label{fig6}
\end{figure}

In order to be able to compute the spectral function, we require all
the eigenstates of the coupled system.
This sets another limit of the basis size, which, by necessity, was
chosen to be about $11000$ basis states: $1000$ of the two-pair, and
$10000$ of the single-pair configurations.
As a result, the eigenstates of the coupled system, defined in
Eq.~(\ref{mixed_state}), can be written as
\begin{equation}
|K\rangle = \sum_{i,j,\alpha,\beta}
C^{K}_{i,j,\alpha,\beta} |i,j,\alpha,\beta\rangle
+ \sum_{i,\alpha}
D^{K}_{i,\alpha} |i,\alpha\rangle.
\end{equation}
while the pure XX states are described by Eq.~(\ref{correlated_biexciton}).
Since the spectral function is defined in terms of a projection of one
of the XX states onto the spectrum $|K\rangle$, in this case we have:
\begin{equation}
A_{p,XX}(K) = |\left._1\langle\right. XX|K\rangle|^2 
= \left|  \sum_{i,j,\alpha,\beta} \left( A^{p,XX}_{i,j,\alpha,\beta} \right)^*
C^{K}_{i,j,\alpha,\beta}   \right|^2.
\end{equation}
In Fig.~\ref{fig6} we plot the spectral function $A_{1,XX}(K)$ of the
XX ground state ($p=1$) as a function of the energy of the eigenstates
of the coupled system.
The red vertical lines represent the XX (i.e., not mixed with
$X_1^{*}$) eigenstates, 
while the gray vertical lines denote the energies $\varepsilon(K)$ of
the mixed eigenstates $|K\rangle$.
The values of the spectral functions for different $K$ are represented
by the height of the blue bars.
We find that the value of the spectral function is $0.847$ at the
energy corresponding to that of the unmixed XX ground state, 
and smaller by about an order of magnitude at the energy corresponding
to the next state $|K\rangle$.
Its value falls off very fast as we move away from this energy, becoming
negligibly small already at about $1$ meV away.
We conclude that our XX level is coupled very weakly to the underlying
spectrum of the $X_1^*$ states and is not  resonant with any of them.

\begin{figure}
\includegraphics[width=0.4\textwidth]{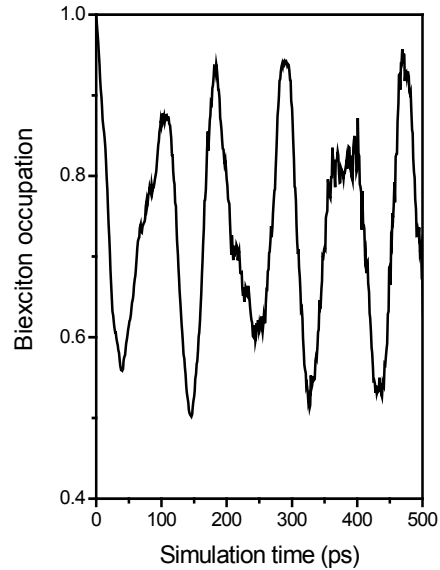}
\caption{(Color online)
  Time evolution of the probability of finding the coupled
  bi-exciton-exciton system in the ground bi-exciton state. }
\label{fig7}
\end{figure}

Next we examine how this weak coupling translates into the dynamics of
the coupled system.
We assume that the system is prepared in the XX ground state and
simulate its time evolution as described in
Section~\ref{spectral_theory}.
The resulting probability of finding the system in the state
$|XX\rangle_1$ at time $t$ is plotted in Fig.~\ref{fig7}.
We find that this probability oscillates around the value of
$|A_{1,XX,max}|^2=0.72$ with no appreciable decay.
The oscillations are not regular, as there are several states which
are effectively coupled to $|XX\rangle_1$ and the values of coupling
elements vary strongly from state to state.
It is clear, however, that we cannot define the XX lifetime
in our system in a meaningful fashion.

\subsection{Bi-exciton decay and lifetime}

The treatment of our system in terms of the coherent evolution of
weakly coupled multiexciton states does not reflect the experimental
situation, in which finite lifetimes, of order of tens of
picoseconds, are found.\cite{schaller_pietryga_nanolett07,ellingson_beard_nanolett05,schaller_agranovich_natphys2005,califano_acsnano2009} 
In Section~\ref{spectral_theory} we have demonstrated that we can
obtain a population of the multiexciton state which decays in time
even within the coherent model if its spectral function is
sufficiently broad. 
In our system, such broadening can occur as a result of mixing of the
$X_1^*$ one-pair configurations, which can be directly coupled to the
two-pair configurations, and the $X_2^*$ one pair configurations,
which exhibit no such coupling.
To account for this mixing, however, we would have to generate {\em
  all} the single-pair configurations with energies falling into the
chosen energy window.
This results in prohibitively large basis sizes.
The second broadening mechanism involves relaxation of
the one-pair configurations due to phonons.
Such relaxation is not relevant for the XX subsystem, as we focus
on the XX ground state only.
Coherent simulations of the time evolution of such a system have been
reported,\cite{witzel_shabaev_prl2010} however there the electronic
structure was computed within the four-band $k\cdot p$ approach.
Our simulations on model systems indicated that if the relaxation
times of the single-pair states are very short, the dynamics of
the biexciton can be described by a single lifetime obtained using the
Fermi's Golden
Rule\cite{shabaev_efros_nanolett06,schaller_agranovich_natphys2005,allan_delerue_prb2006,allan_delerue_prb2008,rabani_baer_nanolett08,franceschetti_an_nanolett06} 
\begin{equation}
{1\over\tau_{1}} = {2\pi\over\hbar}
\sum_{i} |\left._1\langle\right. XX| V_{ee} | X\rangle_i|^2
\delta(E_{1,XX} - E_{i,X}),
\end{equation}
where $E_{1,XX}$ is the ground state energy of the biexciton state and
$E_{i,X}$ is the energy of the $i$-th exciton state.
The influence of scattering and relaxation processes is introduced
into this rule by broadening the delta function such that
$\delta(E_{1,XX} - E_{i,X}) \rightarrow (\Gamma/2\pi) / [ (E_{1,XX} -
E_{i,X})^2 + (\Gamma/2)^2]$, where $\Gamma$ is a model broadening.
In an attempt to model fast relaxation processes, one typically
chooses $\Gamma$ to be sufficiently large so that the result of the
calculation does not depend on it. 

To make contact with previous
work,\cite{shabaev_efros_nanolett06,schaller_agranovich_natphys2005,allan_delerue_prb2006,allan_delerue_prb2008,rabani_baer_nanolett08,franceschetti_an_nanolett06}
we begin by assuming that the states $|XX\rangle_1$ and $|X\rangle_i$
are uncorrelated and are simply represented by the two-pair and
one-pair configurations, respectively.
The single-pair energies are computed as the respective expectation
values of the quasi-particle Hamiltonian 
(\ref{hamiltonian_quasiparticle}): 
$E_{i,X} = \left._i\langle\right.X|\hat{H}_{QP}|X\rangle_i$.
On the other hand, we treat the XX energy $E_{1,XX}$ as an independent
variable, whose value will be tuned artificially in order to examine
the values of lifetimes over an energy region.
However, to compute the matrix element we always use the XX ground
state.

\begin{figure}
\includegraphics[width=0.4\textwidth]{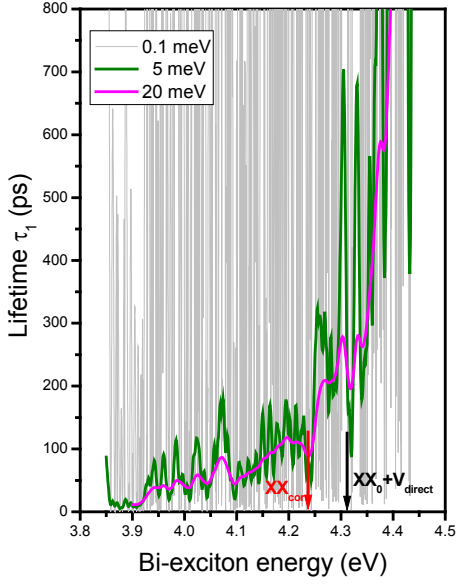}
\caption{(Color online)
 Lifetime of the bi-exciton in the coupled bi-exciton-exciton
  system computed using the Fermi's Golden Rule with the broadening
  $\Gamma=0.1$ meV (gray), $5$ meV (green), and $20$
  meV (magenta). Both exciton and bi-exciton states are uncorrelated.}
\label{fig8}
\end{figure}

In Fig~\ref{fig8} we show the lifetimes computed as a function of the
biexciton energy for four broadenings $\Gamma$.
The actual ground-state energy of the uncorrelated XX is marked with
the black arrow, while the energy corresponding to the fully
correlated XX is denoted by the red arrow.
We find that for a small broadening of $\Gamma=0.1$ meV, the lifetime
as a function of the XX energy exhibits very fast oscillations over
many orders of magnitude. 
Increasing the broadening leads to averaging of these oscillations and
an emergence of a monotonic dependence of $\tau_1$ on $E_{1,XX}$.
This dependence seems to be converged for $\Gamma=20$ meV.
We find that the lifetime increases with the energy, which is
consistent with the decrease of the coupling element with energy (see
Fig.~\ref{fig4}).
In general, we find that depending on the biexciton energy the
calculated lifetimes can vary from about $50$ ps to about $800$ ps.
As we change the XX energy from its uncorrelated to
correlated value, the lifetime changes by a factor of $3$.
It is clear, therefore, that accounting for correlations in the
many-body states is of crucial importance.

\begin{figure}
\includegraphics[width=0.4\textwidth]{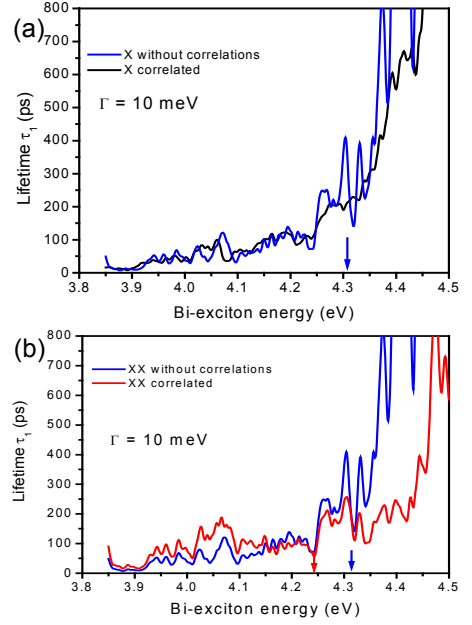}
\caption{(Color online)
 Lifetime of the bi-exciton in the coupled bi-exciton-exciton
  system computed using the Fermi's Golden Rule assuming that the
  exciton is correlated and bi-exciton is not (a) and conversely (b).}
\label{fig9}
\end{figure}

Finally, we examine how the XX lifetime changes if we also include
correlations in the many-body states, i.e., depending upon how we
compute the coupling matrix element in the Fermi's Golden Rule.
Figure~\ref{fig9} shows the lifetime as a function of energy assuming
that the X states are correlated while the XX state is not (a) and the
other way around (b).
Preparation of the X states in their correlated form does not
introduce any drastic changes into the lifetime. 
These correlations can be modeled essentially by an additional
broadening in the density of states of single-pair configurations.
On the other hand, introducing correlations into the XX ground state
increases the lifetime by about a factor of $2$ for lower energies,
while for higher energies the lifetime is reduced by a factor of
$5$.
Clearly, the XX lifetime strongly depends on the details of the sample
and its coupling with relaxation mechanisms.
Detailed {\em ab initio} studies of the nanocrystal electronic
structure and coupling with phonons are necessary to obtain 
predictions of the lifetime which could be compared to the
experimental data.

\section{Conclusions and outlook}

In conclusion, we have presented a theory of the coupling
between states with different numbers of electron-hole excitations in
semiconductor nanostructures.
We have derived the appropriate quasiparticle Hamiltonian
and demonstrated that the states are coupled by a Coulomb interaction
element. 
This element accounts for the process whereby one electron-hole pair
is annihilated and another carrier is scattered within the same band.
We have shown that these processes involve the creation or 
collapse of a charged trion, while the remaining particles act as
spectators. 
The general Hamiltonian for such a coupled system can be diagonalized
exactly in the basis of states with different number of excitations.
The spectral function of the state with higher number of pairs among the
states with fewer pairs was calculated and shown to be related to
the time-dependent evolution of the system by the Fourier transform.

This methodology was applied to analyzing the dynamics of the
bi-exciton (XX) state immersed in the excited exciton (X) states of
similar energy confined in a CdSe spherical nanocrystal.
The NC single-particle energies and states were computed within the
$sp^3d^5s*$ tight-binding approach and the Coulomb scattering matrix
elements were obtained with these atomistically resolved wave
functions.
We derived the spectral function of the XX ground state immersed in
the excited exciton states for an example nanoscrystal of $3.8$ nm
diameter.
We found that the ground XX state is coupled only to a few $X^*$
states close to it in energy.
This property was reflected in the time evolution of the state of the
coupled system, which involved clear coherent oscillations in the
population of the XX state without decay.
Lifetime of the XX state could be defined only by introducing
additional coupling of excited exciton states and their nonradiative
decay mediated by phonons.
The exciton-phonon coupling was accounted for in a model fashion
by broadening the resonance condition in the Fermi's Golden Rule
treatment of the XX lifetime. 
We found that the
lifetime is very sensitive to that broadening (characteristic
relaxation times), if the broadening itself is small, but has a
convergent behavior for large broadenings.
On the other hand, a large change in the XX lifetime
was seen when the correlated character of the XX
state was accounted for.
To investigate this dependence further,  we plan to improve on
the elements of our approach which were carried out in a model
fashion: treatment of surface, a simplified treatment of the
distance-dependent dielectric function, and the approximate treatment
of relaxation mechanisms of the excited exciton states.
Such an approach could be used to provide more
quantitative estimates of XX lifetime in NCs.

\section*{Acknowledgment}
The Authors are grateful to the NRC-NSERC-BDC Nanotechnology Project
and the NRC-CNRS grant for financial support.


\end{document}